\documentclass[reprint,showpacs,preprintnumbers,superscriptaddress,pra,aps]{revtex4-1}

\usepackage{amsmath}
\usepackage{amssymb}
\usepackage{graphicx}
\usepackage{dcolumn}
\usepackage{bm}
\usepackage[dvipsnames]{xcolor}
\usepackage[final]{changes}
\usepackage{lineno}
\usepackage{ulem}
\usepackage{soul}

\makeatletter
\def\Ddots{\mathinner{\mkern1mu\raise\p@	
		box{\kern7\p@\hbox{.}}\mkern2mu
		\raise4\p@\hbox{.}\mkern2mu\raise7\p@\hbox{.}\mkern1mu}}
\makeatother
\newcommand{\be}{\begin{equation}}
\newcommand{\ee}{\end{equation}}

\newcommand{\fig}[1]{Fig.~\ref{#1}}
\newcommand{\figs}[2]{Figs.~\ref{#1} and \ref{#2}}
\newcommand{\Fig}[1]{Figure~\ref{#1}}

\newcommand{\eq}[1]{Eq.~\eqref{#1}}

\newcommand{\bea}{\begin{eqnarray}}
\newcommand{\eea}{\end{eqnarray}}
 
\newcommand{\ba}{\begin{array}}
\newcommand{\ea}{\end{array}}

\newcommand{\bl}{\begin{flalign}}
\newcommand{\enl}{\end{flalign}}

\begin{document}
	\author{Tobias Boolakee}
	\thanks{These authors contributed equally to this work.}
	\affiliation{Department of Physics, Friedrich-Alexander-Universität Erlangen-Nürnberg (FAU), Staudtstrasse 1, D-91058 Erlangen, Germany}
	\author{Christian Heide} 
	\thanks{These authors contributed equally to this work.}
	\affiliation{Department of Physics, Friedrich-Alexander-Universität Erlangen-Nürnberg (FAU), Staudtstrasse 1, D-91058 Erlangen, Germany}
	\affiliation{Stanford PULSE Institute, SLAC National Accelerator Laboratory, Menlo Park, California 94025, USA}
    \author{Antonio Garz\'on-Ram\'irez}
	\affiliation{Department of Chemistry, University of Rochester, Rochester, New York 14627, USA}
    \affiliation{Department of Chemistry, McGill University,
    Montreal, Quebec H3A 0B8, Canada}
\author{Heiko B. Weber}
\affiliation{Department of Physics, Friedrich-Alexander-Universität Erlangen-Nürnberg (FAU), Staudtstrasse 1, D-91058 Erlangen, Germany}
\author{Ignacio Franco}
\affiliation{Department of Chemistry, University of Rochester, Rochester, New York 14627, USA}
\affiliation{Department of Physics, University of Rochester, Rochester, New York 14627, USA}
\author{Peter Hommelhoff}
\affiliation{Department of Physics, Friedrich-Alexander-Universität Erlangen-Nürnberg (FAU), Staudtstrasse 1, D-91058 Erlangen, Germany}
	
\title{Light-field control of real and virtual charge carriers}
\maketitle

\begin{bfseries}
	Light-driven electronic excitation is a cornerstone for energy and information transfer. In the interaction of intense and ultrafast light fields with solids, electrons may be excited irreversibly, or transiently during illumination only. 
	As the transient electron population cannot be observed after the light pulse is gone it is referred to as virtual, while the population remaining excited is called real~\cite{Yablonovitch1989,Yamanishi1987,Schultze2012,Sommer2016}. Virtual charge carriers have recently been associated with high-harmonic generation and transient absorption~\cite{Lucchini2016,Schlaepfer2018,Juergens2020,Sanari2020}, while photocurrent generation may stem from real as well as virtual charge carriers~\cite{Schiffrin2012,Higuchi2017,Chen2018,Langer2020,GarzonRamirez2020,Hanus2021}. 
	Yet, a link between the carrier types in their generation and importance for observables up to technological relevance is missing. 
	Here we show that real and virtual carriers can be excited and disentangled in the optical generation of currents in a gold-graphene-gold heterostructure using few-cycle laser pulses. 
	Depending on the waveform used for photoexcitation, real carriers receive net momentum and propagate to the gold electrodes, while virtual carriers generate a polarization response read out at the gold-graphene interfaces. Based on these insights, we further demonstrate a proof of concept of a logic gate for future lightwave electronics.
	Our results offer a direct means to monitor and excite real and virtual charge carriers. Individual control over each type will dramatically increase the integrated circuit design space and bring closer to reality petahertz signal processing~\cite{Krausz2014,Markov2014}.
\end{bfseries}

Advances in laser technology propelled ultrafast strong-field manipulation of electrons in solids~\cite{Kruchinin2018}. This enabled the injection of charge carriers in large-band gap dielectrics where the potential of virtual carriers for highly reversible electronic switching at optical frequencies has been demonstrated~\cite{Schultze2012,Schiffrin2012,Sommer2016,Lucchini2016,Chen2018}. More recently, the investigation of semiconductors and Dirac materials has relaxed the requirements on lasers for transient charge control and, furthermore, addresses spin, valley and topological control~\cite{Higuchi2017,Ma2017,Schlaepfer2018,JimenezGalan2020,Bai2020}. In these materials the interplay of real and virtual carriers becomes increasingly important and was obscured until now. Using these insights, we can here demonstrate that the combined excitation of both carrier types in graphene brings light-field-driven logic switching in reach.

\begin{figure*}[t!] 
	\begin{center}
		\includegraphics[width=11cm]{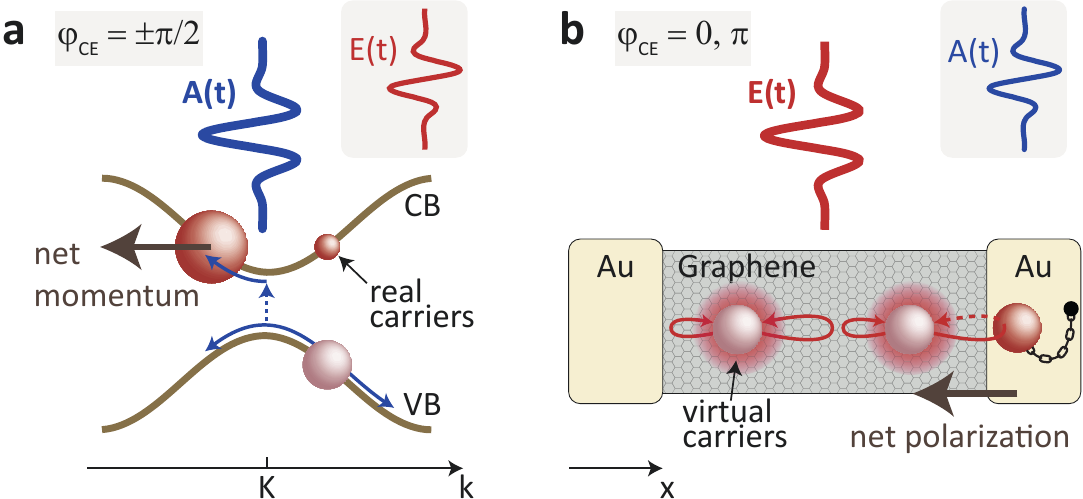}
		\caption{\textbf{Role of virtual and real charge carriers in current generation.} \textbf{a,} For $\varphi_\text{CE} = \pm\pi/2$ and by virtue of the nonlinearity of the interband transition, valence band electrons are excited to the conduction band, and the different magnitudes for positive and negative field amplitudes of the vector potential $A(t)$ (blue waveform) break the residual population symmetry (red spheres) in momentum space, resulting in net momentum after the laser pulse is gone. \textbf{b,} In turn, for $\varphi_\text{CE} = [0, \pi]$, $E(t)$ (red waveform) has different magnitudes for positive and negative field amplitudes and, thus, generates asymmetry in the real-space electron distribution with respect to the gold-graphene interfaces during the pulse. This asymmetry leads to a net polarization, which is probed via the gold electrodes that capture transiently deflected carriers, as indicated by the chained sphere. We assign residual currents that peak at $\varphi_\text{CE} = \pm \pi/2$ as arising from real carriers and currents that peak at $\varphi_\text{CE} = [0, \pi]$ from virtual carriers. Thus, in the experiment, the CEP dependence allows a decomposition of the photo-response into real and virtual carrier contributions.}
		\label{Fig_1}
	\end{center}
\end{figure*}

We use a symmetric heterostructure with graphene as a photoactive material coupled to two gold electrodes (\fig{Fig_1}\textbf{b}) to disentangle real and virtual carriers in the optical generation of currents. As shown in \fig{Fig_1}\textbf{a}, a strong laser pulse acting on the graphene coherently drives intraband motion of electrons in their particular bands at optical frequencies (solid blue arrows) and simultaneously excites electrons from the valence to the conduction band (dashed blue arrow). The intraband dynamics is solely determined by the shape of the vector potential $A(t)=-\int_{-\infty}^t E(t’)dt’$ (blue waveform) associated with the driving optical field $E(t)$ (red waveform)~\cite{Kruchinin2018}. Interband transitions in the form of nonlinear Landau-Zener tunneling events are strongly enhanced close to the K-points~\cite{Ishikawa2010,Heide2021}. Therefore, the interband excitation depends sensitively on the trajectory an electron undergoes. In their excitation, real and virtual carriers are not distinguished by the laser pulse. However, depending on the symmetry of the driving waveform, the two types of carriers may result in a net momentum after the pulse is gone or a net polarization during the pulse. 

Real charge carriers contribute to a residual current if the light field imprints a net momentum to them. Hence, we consider the interaction in momentum space. Net momentum is maximized when the vector potential $A(t)$ has the largest difference in magnitude for positive and negative amplitudes (\fig{Fig_1}\textbf{a}, blue waveform). Carriers starting left and right of the K-points experience different excitation probability, leading to an asymmetric band population (\fig{Fig_1}\textbf{a}, red spheres)~\cite{Heide2021}. The corresponding $E(t)$ is anti-symmetric (\fig{Fig_1}\textbf{a}, red waveform) resulting in a carrier-envelope phase (CEP) of the pulse of $\varphi_\text{CE}=\pm\pi/2$ (Extended Data \fig{Fig_S6}). Real carriers with net momentum travel through the graphene and generate a measurable residual current. For a CEP of 0 and $\pi$, the vector potential is anti-symmetric, thus no net momentum is injected.

By contrast, virtual charge carriers can also contribute to a net current by generating a net polarization. While virtual charge carriers disappear after the light-matter interaction, they can still be detected as a net current, provided that their distinctive behavior is not substantially changed by the measurement. Here, they are probed at the graphene-gold interfaces where transiently displaced charges are separated and rectified to form a net current during the laser pulse (\fig{Fig_1}\textbf{b})~\cite{GarzonRamirez2020}. As supported by our simulations (\fig{Fig_3}), the electrode interfaces serve as an ideal probe to detect virtual carriers electrically. Those carriers that are captured lose their virtual nature and become real (\fig{Fig_1}\textbf{b}, chained red sphere).  As these carriers are localized in the vicinity of the electrodes, a real-space representation is appropriate. Net polarization is maximized when the driving optical field $E(t)$ is maximized for one half-cyce, i.e., for $\varphi_\text{CE}=[0,\,\pi]$ (\fig{Fig_1}\textbf{b}, red waveform). For an anti-symmetric optical field ($\varphi_\text{CE}=\pm\pi/2$), the charge carriers experience equal trajectories towards both electrodes (left and right) and consequently the net polarization is zero.

Strikingly, the two contributions, real carriers with net momentum and virtual carriers with net polarization, are maximized for orthogonal carrier-envelope phases shifted by $\pi/2$, as net momentum is governed by a waveform symmetry of the vector potential, whereas net polarization is determined by a corresponding symmetry in the electric field. Therefore, the measured current response for different CEPs serves as a smoking gun experiment to identify and control the impact of real and virtual carriers in the current generation process. 

To experimentally disentangle real and virtual carriers we thus measure the amplitude and phase of the residual CEP-dependent current induced by CEP-stable 6\,fs near-infrared laser pulses in monolayer graphene contacted to two gold electrodes using a dual-phase lock-in scheme. With the lock-in scheme, the CEP-dependent current is isolated from a CEP-independent photocurrent background. Importantly, graphene itself exhibits a CEP-dependent current response, which we use as an absolute gauge of the CEP in our measurements~\cite{Higuchi2017,Ishikawa2010}. We focus the laser pulses tightly (1.8 $\mu$m 1/e$^2$ intensity radius) to the center of the structure to avoid any spatial symmetry breaking. Details on setup and measurement are given in the Methods and Extended Data \figs{Fig_S2}{Fig_S3}.

\begin{figure*}
	\begin{center}
		\includegraphics[width=15cm]{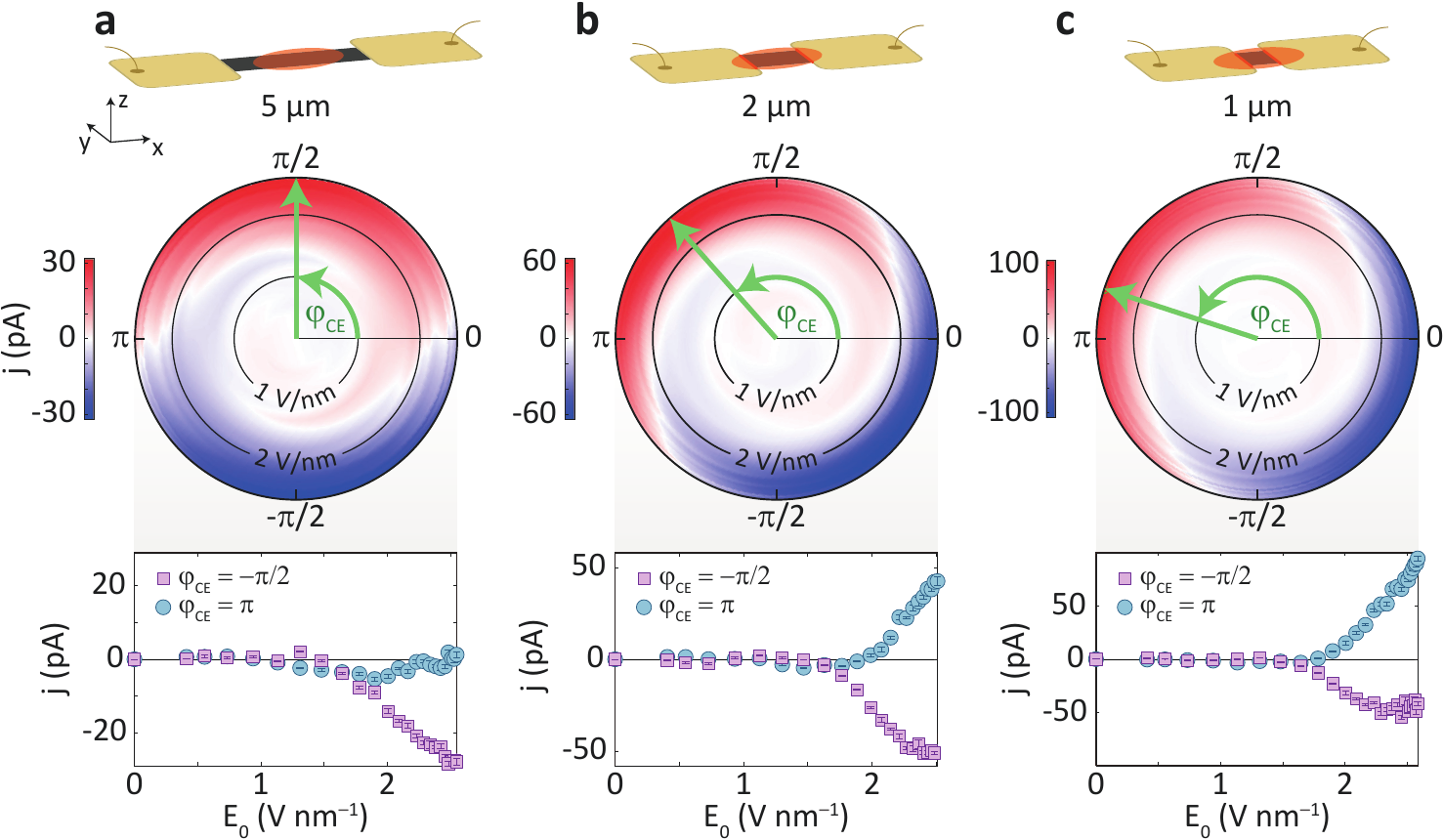}
		\caption{\textbf{Measured phase-resolved CEP-dependent current for graphene strip lengths $\boldsymbol{L=5,2,1}$\,$\boldsymbol{\mu}$m.} The gold-graphene-gold heterostructures shown at the top are illuminated with a Gaussian laser spot (1.8\,$\mu$m 1/e$^2$ intensity radius) placed in the center of each structure and indicated by the red circles. Below, the radius of the polar plots corresponds to $E_0$, which increases from 0 to 2.5\,V/nm, while the polar angle encodes the CEP $\varphi_\text{CE}$. The color coding represents the measured CEP-dependent current. \textbf{a,} The contact electrodes separated with $L=5$\,$\mu$m are hardly illuminated. \textbf{b, c,} Decreasing the graphene strip length to $L=2$\,$\mu$m (\textbf{b}) and $L=1$\,$\mu$m (\textbf{c}) results in an additional current contribution that peaks at $\varphi_\text{CE} = 0, \pi$. The green arrows point towards the maximum current. The bottom plots show the currents projected along the [0, $\pi$]-axis and $[\pm\pi/2]$-axis as a function of $E_0$. Error bars indicate the standard deviation.}
		\label{Fig_2}
	\end{center}
\end{figure*}

\begin{figure}
	\begin{center}
		\includegraphics[width=7.5cm]{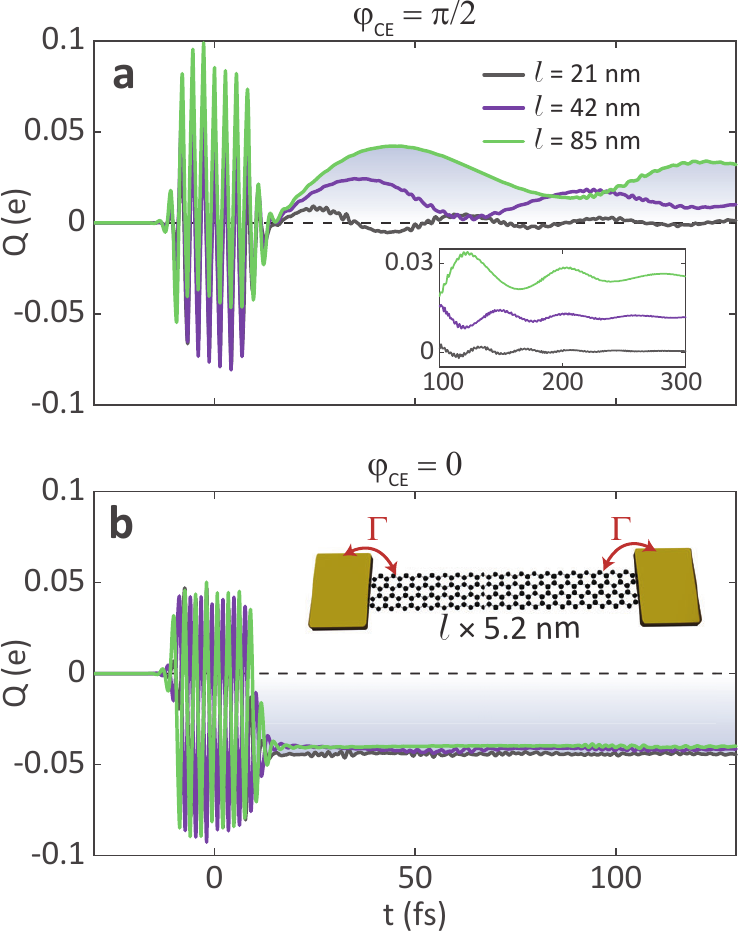}
		\caption{\textbf{Charge transfer simulations for gold-graphene-gold heterostructures (TD-NEGF).} \textbf{a, b,} Charge transfer dynamics for a CEP of $\varphi_\text{CE}=\pi/2$ (\textbf{a}) and $\varphi_\text{CE}=0$ (\textbf{b}) in graphene strips of different length, as indicated. The laser pulse with $E_0=2.3$~V/nm is centered at $t=0$. Inset (\textbf{a}), extended time evolution revealing the net charge transfer after momentum relaxation. Inset (\textbf{b}), schematic of the heterostructure with length $l$ (21, 42, and 85~nm) and a fixed width of 5.2\,nm. Charge transfer between graphene and the electrodes is modeled with a rate $\Gamma/\hbar$ with $\Gamma=0.1$\,eV.}
		\label{Fig_3}
	\end{center}
\end{figure}

\Fig{Fig_2} shows the full phase and amplitude information of the measured CEP-dependent current for three different graphene strip lengths of $L=5,2,1$\,$\mu$m, all 1.8\,$\mu$m wide. We choose these particular strip lengths because they allow us to investigate the role of the gold electrodes in probing the virtual carriers: For $L=5$\,$\mu$m, the laser focus hardly illuminates the gold electrodes and, thus, the role of interface-prone virtual carriers in the current generation process becomes less important. For $L=1$\,$\mu$m, the graphene-gold interfaces are close to the point of maximum field strength, hence the contribution of the virtual carriers may be expected to be dominant (\fig{Fig_2}\textbf{a} vs. \textbf{c}). $L=2$\,$\mu$m lies between the two extreme cases (\fig{Fig_2}\textbf{b}). In the polar plots, the radius corresponds to the peak optical field strength $E_0$, increasing up to 2.5\,V/nm, and the angle encodes the CEP. The bottom panels show line-outs of the currents for a CEP of $-\pi/2$ and $\pi$. 

In \fig{Fig_2}\textbf{a}, the case of the 5\,$\mu$m graphene strip, a carrier-envelope phase of $\pm\pi/2$ induces the largest current response with a strongly nonlinear increase starting to become significant at 1.8\,V/nm. This current is a strong indicator for net momentum injection based on real carriers, as previously reported~\cite{Higuchi2017}. For the maximum applied electric field strength of $E_0=2.5$\,V/nm the current reaches $27\pm2$\,pA. In contrast, no net current is driven with a CEP of 0 and $\pi$, even for the highest field strengths. We note that for longer graphene strips ($L>10$\,$\mu$m) this phase dependence remains unchanged~\cite{Boolakee2020}.

Reducing the distance of the metal electrodes to 2\,$\mu$m (\fig{Fig_2}\textbf{b}) and 1\,$\mu$m (\fig{Fig_2}\textbf{c}) results in illumination of the gold-graphene interfaces. We observe that the green arrows in the polar plots, pointing towards the maximum current, rotate counterclockwise towards $\varphi_\text{CE} = \pi$. In the projections for the 2\,$\mu$m strip (\fig{Fig_2}\textbf{b}), the current driven by $\varphi_\text{CE}=\pi/2$ reaches a maximum value of $51\pm2$\,pA for 2.5\,V/nm. Thus, in the projections, strikingly, a second current contribution emerges for $\varphi_\text{CE}=0$ and reaches a similar value of $43\pm5$\,pA. For a graphene strip with $L = 1$\,$\mu$m (\fig{Fig_2}\textbf{c}) the $\varphi_\text{CE}=0$ current contribution even dominates over the $\pi/2$-response. The $\pi/2$-current again settles at $42\pm4$\,pA, while the current driven by $\varphi_\text{CE}=0$ strongly increases to $94\pm4$\,pA. We find, thus, that the current is maximum for $\varphi_\text{CE}\approx0.9\pi$ (\fig{Fig_2}\textbf{c}, green arrow).

To support the dominating role of virtual carriers for small $L$ and its peaking response for $\varphi_\text{CE}=[0,\,\pi]$, we perform real-space time-dependent non-equilibrium Green's function (TD-NEGF) simulations~\cite{Zhang2013} of the laser-induced electronic transport along metal-graphene-metal nano-structures (see Methods for details). We employ $5.2$\,nm$\times l$ graphene strips of varying length $l = 21, 42, 85$\,nm. The terminal carbon atoms are connected to the metallic contacts and permit charge exchange between the graphene and the metal contacts at a rate $\Gamma/\hbar$, with $\Gamma=0.1$\,eV (\fig{Fig_3}\textbf{b}, inset). Thus, the model explicitly takes into account the crucial role of the metal-graphene interfaces in probing and collecting carriers localized at the graphene strip boundaries. We choose a Gaussian laser pulse with a duration of 6\,fs (FWHM) and central photon energy $\hbar\omega=1.5$\,eV, matching the experimental conditions. The electric field strength is chosen to 2.3\,V/nm, where significant current generation is observed in the experiment, see \fig{Fig_2}.

\Fig{Fig_3} shows the resulting temporal evolution of the charge $Q$ transferred across the heterostructure illuminated with a laser pulse of $\varphi_\text{CE}=\pi/2$ (\fig{Fig_3}\textbf{a}) and $\varphi_\text{CE}=0$ (\fig{Fig_3}\textbf{b}) for the three graphene strip lengths. For $\varphi_\text{CE}=\pi/2$, the charge captured by the electrodes during the laser pulse oscillates with the optical period, as each optical half cycle drives charge carriers alternately into the two electrodes. Right after the pulse, no net charge transfer across the graphene strip is observed. In turn, when the laser pulse is gone, the charge collected by the electrodes starts to increase linearly with time, as electrons and holes, launched with net momentum, reach the electrodes via ballistic and diffusive transport. Importantly, the momentum magnitude, i.e., the initial slope of $Q$, imparted on the charge carriers is independent of the graphene strip length $l$ and is solely governed by the optical waveform. Part of the charge carriers are scattered back at the metal interfaces after several 10s of fs, leading to oscillations in the transferred charge. For a larger system size approaching the experimental one and thereby surpassing the length scale of ballistic transport, we expect these oscillations to disappear. The net charge transfer settles at $Q(t\rightarrow\infty)\ne0$ after the current flow has decayed on a $\sim$100\,fs time scale, consistent with previous literature~\cite{Malic2011,Gierz2013}, see inset of \fig{Fig_3}\textbf{a} and Methods for details. This contribution to the current increases with graphene strip length as the photoactive area increases. 

By contrast, for $\varphi_\text{CE}=0$, right after the transient charge oscillations, the amount of collected charges reaches a constant nonzero value of $Q(t\rightarrow\infty) = -0.04\,e$, in good agreement with the experiment (see Methods for details). This net charge transfer is equal to the polarization build-up of displaced charges during the laser pulse, hence it stems directly from virtual carriers captured at the interfaces. As expected for an interface effect, this contribution is independent of the graphene strip length. 

The transfer of the identified temporal symmetries of the laser pulses to the system allude to the existence of a general principle, akin to those that have been identified for periodic driving~\cite{Franco2008}. Moreover, as the transient charge dynamics occur on a timescale faster than electronic dephasing, the CEP dependence and the direction of macroscopic charge transfer is preserved. A phenomenological dephasing is incorporated in the simulations shown in \fig{Fig_3}, in agreement with previous results on ultrafast carrier dynamics (see Methods for details)~\cite{Gierz2013,Neufeld2021,Floss2019,Heide2021b}.

\begin{figure}[h!]
	\begin{center}
		\includegraphics[width=7.5cm]{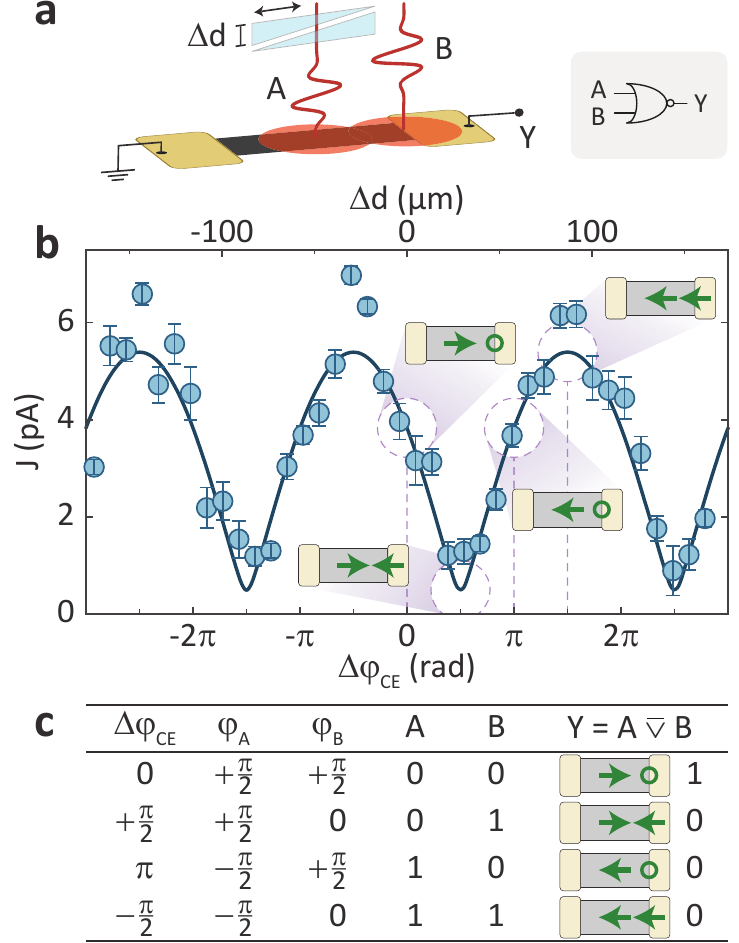}
		\caption{\textbf{Petahertz logic gate.} \textbf{a,} Two laser pulses induce currents via real ($A$, bulk graphene) and virtual ($B$, interface) carriers. Their respective CEPs $\varphi_A$ and $\varphi_B$ are set by a relative phase $\Delta\varphi_\text{CE}$, controlled by the SiO$_2$ thickness $\Delta d$ of a wedge pair passed by pulse $A$; additionally, both CEPs are modulated by a carrier-envelope offset frequency (not shown, see Methods). $\varphi_A$ and $\varphi_B$ determine the resulting current and thus the logic output level $Y$. This way, the structure may operate as a logic NOR gate, for instance (right). \textbf{b,} Measured CEP-dependent current (blue data points) as a function of $\Delta\varphi_\text{CE}$. A model simulation (blue line) almost perfectly matches the obtained currents. The four insets depict the net current induced by the two laser pulses as required for the operation of a logic NOR gate: green arrow to the right depicts a positive current, green arrow to the left a negative current, while a green ring means zero current. 
		We note that the current shown here reflects the direct output of the measurement scheme used (dual-phase lock-in amplification, see Methods), i.e, the root mean square value of an alternating current; its particular sign is unveiled in Extended Data \fig{Fig_S9}. Error bars indicate the standard deviation. \textbf{c,} Truth table for operation of a logic NOR gate. We assign a positive current larger than $+2$\,pA to a logic 1 while a smaller (or negative) current is assigned to 0. Hence, the individual current contributions can be chosen such that a sufficiently large positive current is observed in the first case only, representing a NOR gate. Other types of logic gates (e.g., AND, OR, NAND) can be straight-forwardly obtained by changing the values of $\varphi_A$ and $\varphi_B$ (see Extended Data \fig{Fig_S10}).}
		\label{Fig_4}
	\end{center}
\end{figure}

In agreement with the simulations, for the large graphene strip length ($L=5$\,$\mu$m) the contribution of real carriers dominates the CEP-dependent current and makes it peak at $\varphi_\text{CE}=\pm\pi/2$. As the strip length is decreased and the laser focus illuminates the entire length, the interfaces start to probe virtual carriers efficiently, leading to a transition of the maximum current to $\varphi_\text{CE}=[0,\pi]$. Thus, the simulations fully explain our experimentally observed phase shifts. Further numerical analysis of the population dynamics in graphene identifying the distinct roles of real and virtual carriers in the charge transfer is given in the Methods.

The magnitude of the observed $\pi/2$ current is determined by the photoactive area in the graphene and is influenced by phonon and impurity scattering of the charge carriers during propagation to the electrodes~\cite{Gierz2013,Tetienne2017}. The larger the photoactive area, the larger the current (\fig{Fig_3}\textbf{a}). However, in the experiments, as the graphene strip length increases, the $\pi/2$ current decreases, likely because of charge scattering~\cite{Boolakee2020}. 

Both real and virtual charge carriers are of utmost importance in the optical control of electrons. Excitation of virtual carriers is the key for generating reversible currents on sub-cycle timescales of the laser field~\cite{Schiffrin2012, Krausz2014, Tan2018, Heide2020}. In addition, the generation of real carriers with net momentum is a result of interfering quantum pathways~\cite{Fortier2004,Higuchi2017}. Thus, real carriers may offer a unique platform to exploit the transient quantum mechanical phase evolution of electrons for signal processing before they are probed by the electrodes. Hence, the understanding of both carrier types may be utilized in designing future petahertz circuit architecture, with a much larger design space available than with one carrier type only.

We show this with a concrete example in \fig{Fig_4}\textbf{a}, where we demonstrate the proof of concept for a logic gate relying on real and virtual charge carriers. Pulse $A$, focused to the center of the gold-graphene-gold heterostructure, may inject real carriers that propagate towards the gold-graphene interface illuminated by pulse B. Pulse B is synchronized to the arrival of the current injected by pulse A and may control a second current component by virtual carriers at that interface. The essence of the logic gate is to encode bits into the CEP of the two laser pulses $A$ ($\varphi_A$) and $B$ ($\varphi_B$) and have the system sum up the resulting current contributions. The total current carries the desired logic output $Y$. If the current is larger than $+2$\,pA, we assign it to $A=1$, else $Y=0$.

\Fig{Fig_4}\textbf{b} shows the CEP-dependent current as a function of the relative phase $\Delta\varphi_\text{CE}=\varphi_A-\varphi_B$, which is controlled by changing the thickness $\Delta d$ of SiO$_2$ passed by pulse $A$. A peak optical field strength of 2.3\,V/nm is employed for both pulses. Based on the excellent agreement between the measured current and a model simulation (blue line, see Methods for details) we can infer the individual current components injected by pulse $A$ and $B$ (\fig{Fig_4}\textbf{b}, insets).

We can now pick four scenarios of incident CEPs $\varphi_A$ and $\varphi_B$ as specified in the truth table (\fig{Fig_4}\textbf{c} and \fig{Fig_4}\textbf{b}, insets) to design a NOR gate. In the first case ($\Delta\varphi_\text{CE}=0$, and $\varphi_A=\varphi_B=+\pi/2$, first line in \fig{Fig_4}\textbf{c}), pulse $A$ injects a net current based on real carriers into the right electrode while the generation of net current due to virtual carriers by pulse $B$ is absent. For these CEPs, we obtain $+4$\,pA in the experiment, hence $Y=1$. In the second case ($\Delta\varphi_\text{CE}=+\pi/2$), both pulses induce currents in opposite direction, leading to a cancellation of currents. We find in the experiment a total current of $+1$\,pA only, hence $Y=0$. Third, for $\Delta\varphi_\text{CE}=\pi$ and as in the first case, $\varphi_B$ is chosen not to excite a net current at the interface while $\varphi_A$ is flipped to $-\pi/2$ to inject a net current to the left electrode. The measurement yields $-4$\,pA, so a current, but in opposite direction, hence, again, $Y=0$ (see comment on sign of current in caption to \fig{Fig_4}\textbf{b}). Finally, for $\Delta\varphi_\text{CE}=-\pi/2$ ($\varphi_A=-\pi/2$, $\varphi_B=0$), both laser pulses inject currents along the same (negative) direction, resulting in a total current of $-6$\,pA, hence $Y=0$ again. This series of currents with their respective logic input pairs clearly results in a NOR logic gate with $\varphi_A$ and $\varphi_B$ assigned to logic inputs 0 and 1. 

Ultimately, the CEP may be superseded by trains of successive optical cycles $A$ and $B$ with appropriately shaped waveforms, pushing the bandwidth of the gate to its fundamental limit, the frequency of light~\cite{Garg2016}. Moreover, the use of real carriers may enable efficient switching by utilizing the electron momentum imparted by a single pulse $A$ for several subsequent operations controlled by a sequence of pulses $B$. Hence, energy consumption can be reduced by algorithmic optimization~\cite{Markov2014}. Illumination via nearfield or plasmonic confinement, or the use of optical waveguides, could reduce the gate footprint to the nanometer size, similar to the size of today's transistors~\cite{Krausz2014}; a scale at which our charge transport simulations prove that the mechanisms continue to work.

In summary, we have demonstrated the role of virtual and real charge carriers in the current generation process driven by a few-cycle optical field. They show fundamentally different characteristics in the injection of currents in heterostructures: the injection of net momentum by real carriers in the bulk of the photoactive material and the generation of a net polarization by virtual carriers read out at the interfaces. We expect these insights to offer new degrees of freedom for the implementation of future electronics at optical frequencies, exemplified by a first logic gate introduced and experimentally demonstrated here.

\subsection*{Acknowledgements} 
We thank J. Ristein and M. Hundhausen for discussions. This work has been supported in part by the Deutsche Forschungsgemeinschaft (SFB 953 “Synthetic Carbon Allotropes”, project number 182849149), the PETACom project financed by Future and Emerging Technologies Open H2020 program, ERC Grants NearFieldAtto and AccelOnChip, and the US National Science Foundation under Grant No. CHE-1553939 and CHE-2102386.

\subsection*{Methods}

\setcounter{figure}{0}
\makeatletter 
\renewcommand{\thefigure}{S\@arabic\c@figure}
\makeatother

\subsubsection*{Sample fabrication and characterization.}

\begin{figure}[b!]
	\begin{center}
		\includegraphics[width=7.5cm]{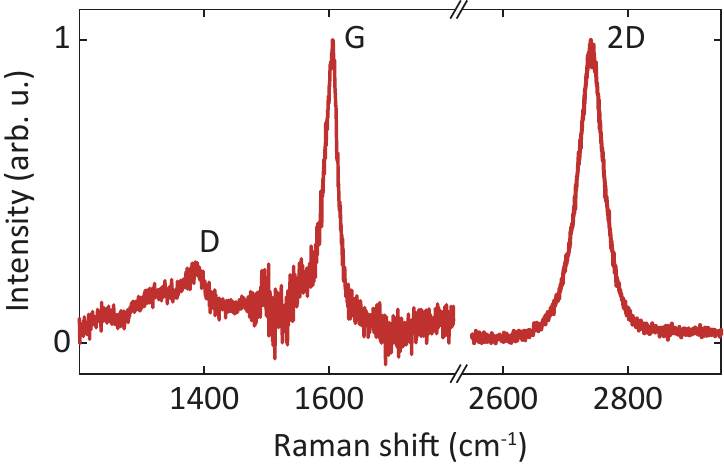}
		\caption{\textbf{Raman spectrum of monolayer epitaxial graphene on SiC.} The coupling to the SiC induces a strain-induced blue-shift of 64\,cm$^{-1}$ to the 2D-peak compared to freestanding graphene, while it consists of one peak only ($\sim$48\,cm$^{-1}$ FWHM) as expected for monolayer graphene~\cite{Roehrl2008}. The occurence of a D-peak indicates defects emerging primarily from domain boundaries~\cite{Emtsev2009}.}
		\label{Fig_S1}
	\end{center}
\end{figure}

Monolayer graphene is epitaxially grown from a 4H-Silicon Carbide (SiC) substrate. Hall measurements yield a free carrier concentration of $n=(8.0\pm0.9)\cdot10^{12}$\,cm$^{-2}$, which implies a Fermi level at $E_\text{F}=0.3$\,eV above the Dirac point, and a mobility of $\mu=860\pm60$\,cm$^2$V$^{-1}$s$^{-1}$ is obtained. Hence, the momentum relaxation time of thermalized carriers can be estimated to 26\,fs from the Drude model. Scanning electron microscopy inspection and Raman spectroscopy confirms the growth of monolayer graphene, see \fig{Fig_S1} for details. Graphene is particularly well suited to carry CEP-dependent currents as it is conducting, has a broadband optical response and avoids optical propagation effects due its 2-dimensional nature~\cite{CastroNeto2009}.  Gold electrodes with a thickness of 30\,nm are deposited with a 5\,nm titanium adhesive layer. Graphene strips with varying length of 1, 2 and 5\,$\mu$m are fabricated via electron beam lithography and subsequent oxygen plasma etching. Electrical contacting from the gold electrodes to a chip carrier is provided via aluminum wire bonding. A scanning electron micrograph of the heterostructures is shown in \fig{Fig_S2}\textbf{a}.

\subsubsection*{Experimental details.}

\begin{figure}
	\begin{center}
		\includegraphics[width=7.5cm]{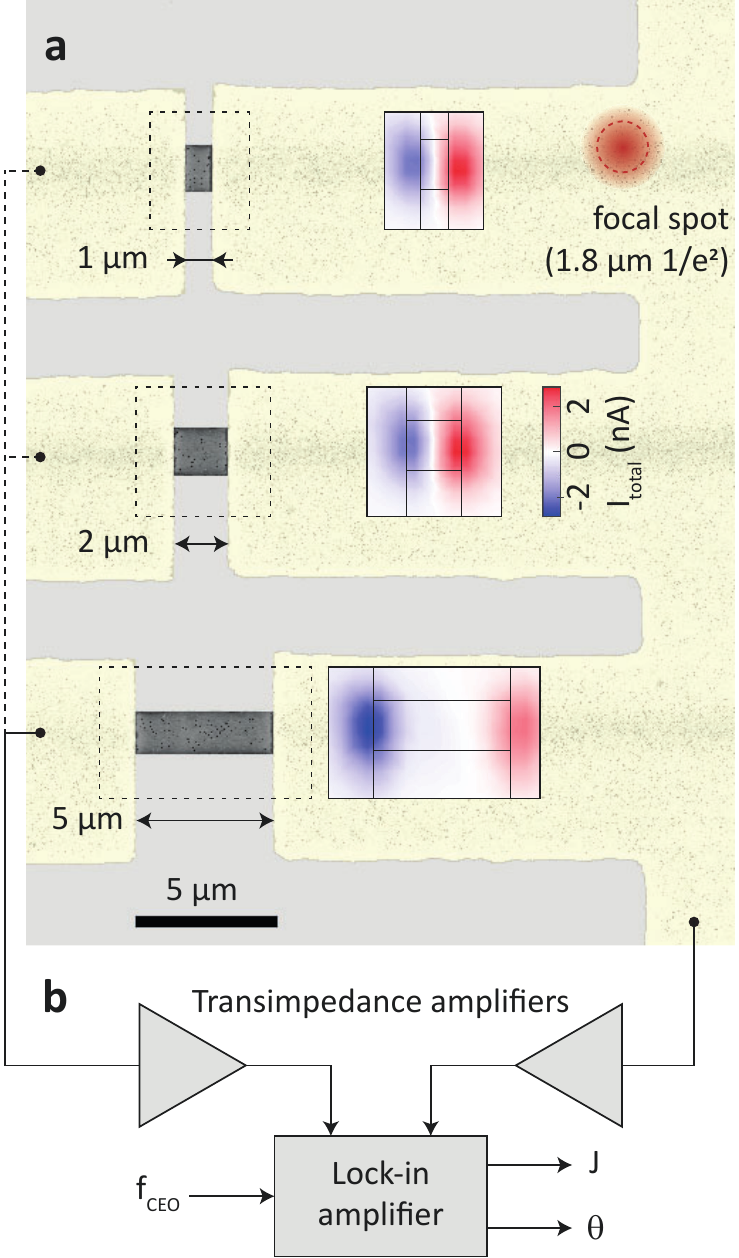}
		\caption{\textbf{Sample and measurement technique.} \textbf{a,} Scanning electron micrograph of the gold-graphene-gold heterostructures with various electrode distances. The yellow coloring indicates the gold electrodes. Insets, total photocurrent as a function of the focal position inside the dashed regions ($E_0 = 0.3$\,V/nm). For all measurements and unless otherwise stated, the laser spot was positioned in the center of the graphene strips such that the total photocurrent is 0. \textbf{b,} Schematic diagram of the measurement scheme. The induced residual current is amplified by transimpedance amplifiers and detected by a dual-phase lock-in amplifier.}
		\label{Fig_S2}
	\end{center}
\end{figure}

All measurements are performed under vacuum conditions ($1\times10^{-8}$\,hPa) at room temperature. CEP-stable 6\,fs laser pulses obtained from a titanium:sapphire laser oscillator with a center photon energy of 1.5\,eV and a repetition rate of 80\,MHz are used to drive currents. The polarization of the optical field is parallel to the graphene strip. Currents injected to the electrodes are amplified with two transimpedance amplifiers. The CEP-dependent current is measured with a dual-phase lock-in amplifier at a carrier-envelope-offset frequency of $f_\text{CEO}=3.3$\,kHz, which is used as lock-in reference (\fig{Fig_S2}\textbf{b}). With this technique we have full information about the current amplitude $J$ and phase $\theta$, which equals the carrier-envelope phase of the exciting pulses up to an arbitrary phase offset. 

\begin{figure*}
	\begin{center}
		\includegraphics[width=14cm]{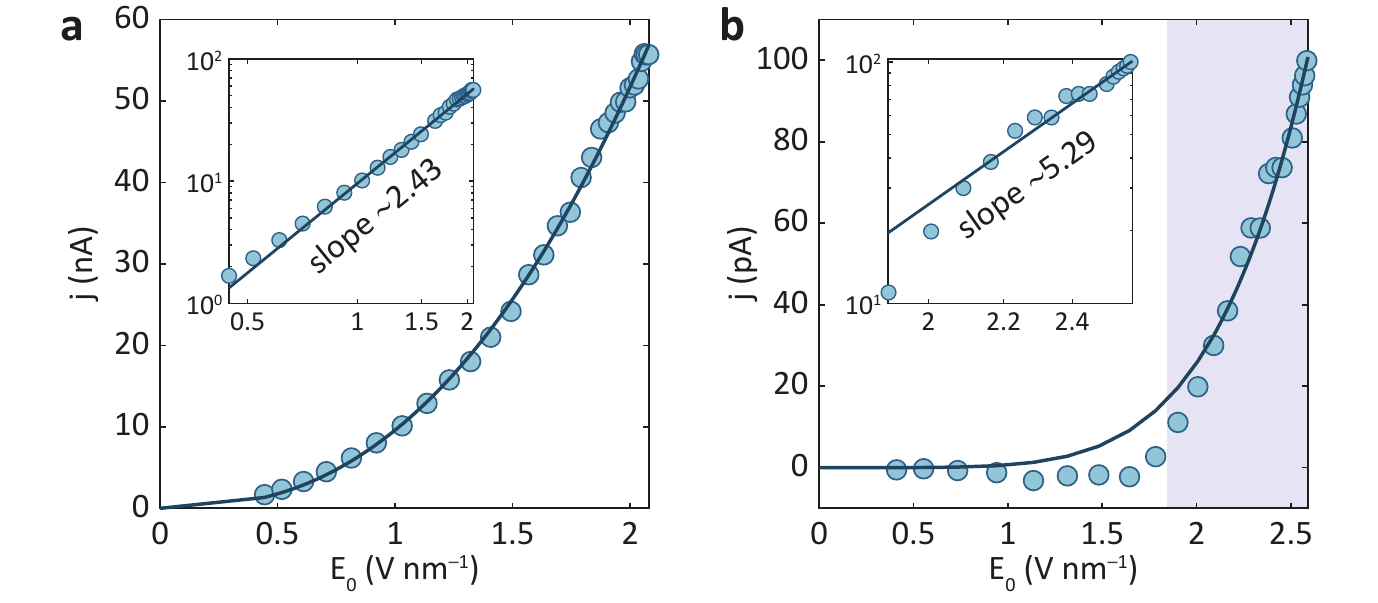}
		\caption{\textbf{Total photocurrent and CEP-dependent current on an $\bm{L=1\,\mu}$m graphene strip as a function of $\bm{E_0}$.} \textbf{a,} Total photocurrent with the laser focus placed at the interface (compare with current maxima in \fig{Fig_S2}a, inset). \textbf{b,} CEP-dependent current for $\varphi_\text{CE}=\pi$ (same data as in Fig.~2\textbf{c}). Insets show the data in a double-logarithmic scale with linear fits, while shaded points only are included in the fit in (\textbf{b}).}
		\label{Fig_S3}
	\end{center}
\end{figure*}

We calibrate this phase offset by measuring the CEP-dependent current on the the $5\,\mu$m graphene strip. Based on symmetry arguments~\cite{Higuchi2017} and numerical simulations~\cite{Kelardeh2015}, the current is maximized for $\theta=\varphi_\text{CE} = \pm\pi/2$. With knowledge on this relation, we are able to disentangle components $j_{\pm\pi/2} = J \cdot \sin(\varphi_\text{CE} )$ and $j_{0, \pi} = J \cdot \cos(\varphi_\text{CE} )$ to determine the current generated at the interfaces via virtual carriers and the bulk current contribution (real carriers). Higher order harmonics of the lock-in reference frequency are not observed in the current, indicating that for all scenarios the phase response of the current follows the CEP linearly.

For all measurements, the laser is focused with an off-axis parabolic mirror to a spot in the center of the graphene strip. This center position is determined by nulling the total photocurrent, which may be generated due to photo-thermoelectric and built-in electrostatic potentials at the metal-graphene interfaces (\fig{Fig_S2}\textbf{a}, insets)~\cite{Gabor2011,Shautsova2018}. We note that the readout of virtual charge carriers is spatially not limited to the atomically sharp gold-graphene interfaces but may extend over the few 100\,nm decay length of the photovoltaic and photo-thermoelectric contact potentials~\cite{Mueller2009,Woessner2016}. In the photocurrent maps of \fig{Fig_S2}\textbf{a}, these potentials become visible in the generation of the total current while the resolution is limited by the focus size. 

Potentially, a CEP-dependent current could result from a small CEP dependence of photo-thermoelectric or photovoltaic currents. To rule this out,  we compare the scaling of the total photocurrent as a function of incident field strength with that of the CEP-dependent current, see \fig{Fig_S3}. Analyzing the currents by linear fits in a double-logarithmic representation (\fig{Fig_S3}, insets) shows that the total current scales approximately quadratically with field strength (i.e., linear with power, cf. with \cite{Shautsova2018}) while the CEP-dependent current scales with a significantly higher nonlinearity of $\sim$5. From the different scaling we conclude that the two current components originate from different mechanisms, i.e., photo-thermoelectric and photovoltaic vs. field-driven.

Each data point shown in Fig.~2 is recorded with 1\,s time constant and integrated for 30\,s; the data points shown in Fig.~4\textbf{b} are integrated for 7\,s. Slow phase fluctuations on the order of seconds are recorded with an out-of-loop $f$-$2f$ interferometer. The phase of the measured current is corrected for slow CEP drifts.

\subsubsection*{Near-field effects.}

\begin{figure}
	\begin{center}
		\includegraphics[width=7.5cm]{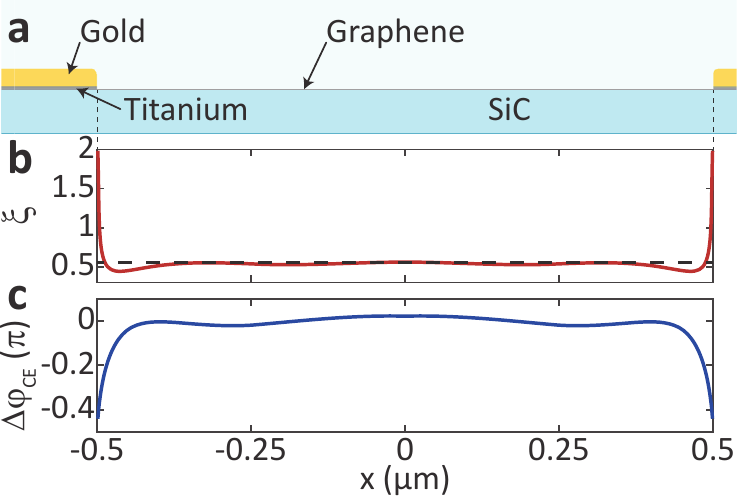}
		\caption{\textbf{Optical near-field simulation of the heterojunction.} \textbf{a,} Schematic of the structure used in FDTD simulations. Two electrodes consisting of 5\,nm titanium and 30\,nm gold separated by a 1\,$\mu$m graphene strip are supported by a SiC substrate. Similar to experimental conditions, the structure is illuminated with a Gaussian focus placed in the center. \textbf{b,} Field enhancement $\xi$ within the graphene layer (red line). In the center, $\xi$ reaches the analytic factor given by the polarization response of bare SiC (gray dashed line) while it rises to 1.4 in the optical near-field of the electrodes. \textbf{c,} Variation of the CEP $\Delta\varphi_\text{CE}$ across the graphene strip. $\Delta\varphi_\text{CE}$ is independent of the exact value of the CEP.}
		\label{Fig_S4}
	\end{center}
\end{figure}

On the surface of the graphene, the vacuum optical field strength is reduced by a factor $2/(1+n_\text{SiC})$ due to the dielectric polarization response of the SiC substrate underneath. $n_\text{SiC}\approx2.6$ is the refractive index of SiC at the laser center wavelength. The electric field strength $E_0$ given in the manuscript includes this factor.

Based on finite difference time domain (FDTD) simulations we identify the impact of optical field distortions in the presence of the gold electrodes. For a structure length of 1\,$\mu$m  the electrodes are illuminated as well and, thus, the optical field on the graphene is enhanced by a factor of up to 2 in the vicinity to the interfaces (\fig{Fig_S4}\textbf{b}). The near-field enhancement decays on a length scale of less than 50\,nm. Simultaneously, we observe that the carrier-envelope phase that arrives at the graphene surface is shifted by $\Delta\varphi_\text{CE}\approx0.4\pi$ in the same range compared to the case without illumination of the electrodes (\fig{Fig_S4}\textbf{c}). 

While our experiments directly rely on the detection of a CEP-response, we note that both the near-field enhancement and the near-field shift in the CEP do not have significant influence on the outcome of our experiments: In the measurements shown in Fig.~2, we observe the onset of a current at $E_0\approx1.8$\,V/nm, regardless of the junction size and the CEP-component. Taking into account the field enhancement by a factor of 2 derived for the 1\,$\mu$m heterostructure (Fig.~2\textbf{c}), the graphene in the vicinity of the electrodes experiences a field strength of 1.8\,V/nm already for $E_0\approx0.9$\,V/nm. In particular for the interface-prone $[0, \pi]$-current a near-field-induced shift of the current onset towards lower field strength is not observed. As we do not observe such a shift in the current onset, we conclude that the collective excitation of the electron density throughout the graphene strip dominates the current response. Consequently, the same applies to the local shift of the CEP, which therefore can be neglected.

\begin{figure}
	\begin{center}
		\includegraphics[width=7.5cm]{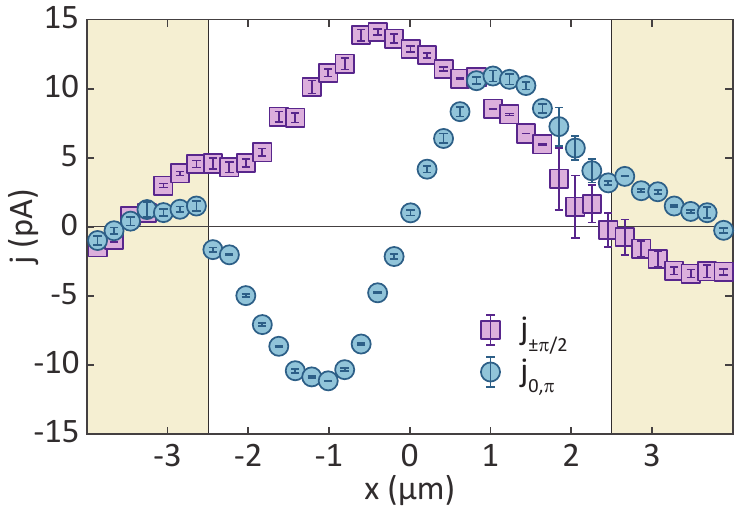}
		\caption{\textbf{Line-scan of CEP-dependent current.} Measured CEP-dependent currents are plotted as a function of the focus position that is moved from one to the other gold electrode (shaded areas) accross a $5\times1.8$\,$\mu$m$^2$ graphene strip on an axis centered to the graphene strip width. The current projections $j_{\pm\pi/2}$ (purple data points) and $j_{0,\pi}$ (blue data points) are shown. A peak field strength of $E_0 = 2.7$\,V/nm is applied. Error bars indicate the standard deviation.}
		\label{Fig_S5}
	\end{center}
\end{figure}

To further rule out the influence of near-field-induced CEP-shifts we measure the CEP-dependent current as a function of the focus position, see \fig{Fig_S5}. The focus is moved accross a $5$\,$\mu$m long graphene strip from one to the other electrode on an axis centered to the graphene strip width. $j_{\pm\pi/2}$ peaks in the center of the graphene strip, and because the electrodes are not illuminated here, we assign this current to real carriers. In contrast, $j_{0,\pi}$ is zero in the center. When the focus position is scanned from one to the other electrode, the current shows peaks, most importantly with current flow to opposite directions. Hence, this current component is sensitive to a real-space broken symmetry in the illuminated region, given here by the gold-graphene interfaces. In accordance with the measurements shown in Fig.~2 we assign $j_{0,\pi}$ to virtual carriers. Note that the small strip lengths shown in Figs.~2\textbf{b} and \textbf{c} yield a current $j_{0,\pi}$ even under illumination in the center because the symmetry breaking with respect to the electrodes is introduced by the waveform only. 

In the following we discuss what we would expect in case of a dominating near-field-induced CEP-shift. First, we assume that the currents originate form real carriers only. While in the center no near-field-induced CEP-shift is observed, real carriers at the electrodes experience a CEP shifted by $\Delta\varphi_\text{CE}\approx0.4\pi\approx\pi/2$. Importantly, the sign of the phase shift is equal at both electrodes. It can therefore explain a current component in $j_{0,\pi}$ due to real carriers but not the different current signs towards the left and right electrode.

Second, we assume that real and virtual charge carriers are excited. With $\Delta\varphi_\text{CE}$ included, we expect currents due to real and virtual carriers to appear at the same lock-in phase. The real carrier current $j_{\pm\pi/2}$ is unaffected while, under the influence of the electrode near-field, the virtual carrier current becomes $j_{0,\pi}=J\cdot\cos(\varphi_\text{CE}+\Delta\varphi_\text{CE})\approx-J\cdot\sin(\varphi_\text{CE})$. Hence, in the lock-in measurement the two types of carriers would be indistinguishable based on their CEP dependence. Again, we conclude that the influence of the near-field on the observed currents is negligible as \fig{Fig_S5} does not reproduce the discussed behavior.

We note that an analysis of the underlying microscopic charge dynamics in the presence of such a spatially varying field is beyond state of the art of computational methods and we therefore conduct all simulations under electric dipole approximation.

To exclude electron emission from the metal electrodes under the presence of the enhanced field as a source for the CEP-dependent current, we remove graphene between the gold electrodes and perform the same measurements. For up to $E_0 = 2.5$\,V/nm, no measurable CEP-dependent current is obtained between two electrodes.

\subsubsection*{Hamiltonian.}

\begin{figure}
	\begin{center}
		\includegraphics[width=7.5cm]{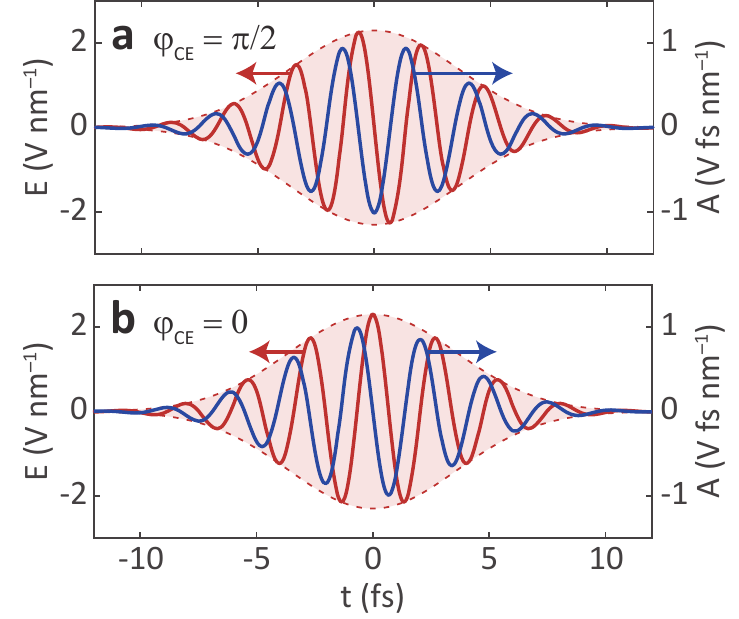}
		\caption{\textbf{Electric field and vector potential of Gaussian laser pulses used for simulations.} The maximum field strength is 2.3\,V/nm, the pulse duration is 6\,fs (intensity FWHM), the center photon energy is 1.5\,eV. \textbf{a,} For a CEP $\varphi_\text{CE}=\pi/2$ the electric field is anti-symmetric with respect to time inversion while the vector potential is symmetric. \textbf{b,} For $\varphi_\text{CE}=0$ the electric field is symmetric and the vector potential is anti-symmetric.}
		\label{Fig_S6}
	\end{center}
\end{figure}

The Hamiltonian for the metal-graphene-metal junction is given by
\begin{equation}
	\label{eq:composite_hamiltonian}
	H(t)=H_{\text{S}}(t)+H_{\text{M}}(t)+H_{\text{SM}}, 
\end{equation} 
where $H_{\text{S}}(t)$ describes the Hamiltonian for graphene, $H_{\text{M}}(t)$ that of the metals, and $H_{\text{SM}}$ the metal-graphene interactions. The graphene is modeled as a zigzag graphene nanoribbon (GNR), 22 carbon atoms wide (5.2\,nm) and 200, 400, or 800 carbon atoms (21, 42, or 85\,nm) long. This nanoribbon is chosen because of its zero band gap and because its density of states is large enough to model bulk behavior. The GNR and its interaction with light in dipole approximation is described by the well-known tight-binding Hamiltonian for Graphene
\begin{equation}
	\begin{split}
		\label{eq:graphenehamil}
		H_{\text{S}}(t)&=\tau\sum_{\langle ij\rangle}(\hat{a}_i^{\dagger}\hat{b}_j+\hat{b}_j^{\dagger}\hat{a}_i)+\\
		&+|e| \sum_i\mathbf{E}(t) \cdot \mathbf{r}_i(\hat{a}_i^{\dagger}\hat{a}_i-1)+\\
		&+|e|\sum_j\mathbf{E}(t) \cdot \mathbf{r}_j(\hat{b}_j^{\dagger}\hat{b}_j-1),
	\end{split}
\end{equation} 
where $\hat{a}_i$ ($\hat{a}_i^{\dagger}$) and $\hat{b}_j$ ($\hat{b}_j^{\dagger}$) are the fermionic annihilation (creation) operators for the two carbon atoms in each unit cell, with tight-binding coupling $\tau=-3.0$\,eV, $\langle ij\rangle$ denotes a sum over nearest-neighbors, and $\mathbf{r}_j$ is the position of C atom $j$. The three nearest neighbors of a given carbon atom are separated by vectors $ \mathbf{d}_1=\frac{a}{\sqrt{3}}\hat{x}$, $ \mathbf{b}_{2,3}=\frac{a}{\sqrt{3}}\left(\frac{1}{2}\hat{x}\pm\frac{\sqrt{3}}{2}\hat{y}\right)$ with lattice constant $a=2.46$\,\AA, where $\hat{x}$ is a unit vector along the direction of junction growth and $\hat{y}$ is a unit vector perpendicular. For the electric field $\mathbf{E}(t)$, a Gaussian pulse as depicted in \fig{Fig_S6} is employed.

The metal contacts are described by the Hamiltonian
$H_{\text{M}}=\sum_{\alpha=\text{L},\text{R}}\sum_q \varepsilon_{\alpha,q}\hat{a}^{\dagger}_{\alpha,q}\hat{a}_{\alpha,q}$, where $\hat{a}^{\dagger}_{\alpha,q}$ and $\hat{a}_{\alpha,q}$ are the fermionic operators for the metal states of energy $\varepsilon_{\alpha,q}$, where $\alpha=\text{L}$ or R denotes the left or right contact, respectively. The metals are taken to be at thermal equilibrium at a temperature of $T=300$\,K and their interaction with the GNR to be well described in the wide band limit (WBL). We suppose that only the unit cells of the edges of the GNR couple to its adjacent metal contact, that is
$H_{\text{SM}}= (\sum_{q\in \text{L}, i'} V_q^\text{L} \hat{a}^{\dagger}_{\text{L},q}(\hat{a}_{i'}+\hat{b}_{i'})
+\sum_{q\in \text{R}, j'} V_q^\text{R} \hat{a}^{\dagger}_{\text{R},q}(\hat{a}_{j'}+\hat{b}_{j'})+\text{H.c.})$, where the label $i'$ ($j'$) runs over carbon atoms in the terminal unit cells adjacent to the left (right) contact and H.c. denotes Hermitian conjugate. The effective coupling between the GNR and the metal contact $\alpha$ is specified by the spectral density $\Gamma_{\alpha}(\varepsilon)=2\pi\sum_q |V_q^{\alpha}|\delta(\varepsilon-\varepsilon_{\alpha,q})$. In WBL the $V_q^{\alpha}$ and the metal density of states $\eta^{\alpha}=\sum_q \delta(\varepsilon-\varepsilon_{\alpha,q})$ are assumed to be energy independent, and therefore $\Gamma_{\alpha}=2\pi|V^{\alpha}|\eta^{\alpha}$ is also energy independent. The quantity $\Gamma/\hbar$ determines the rate of charge exchange between the GNR and the contacts. 

In the simulations we take $\Gamma_{\text{L}}=\Gamma_{\text{R}}=0.1$\,eV, which compares well to the current magnitudes we obtain in the experiment: For $L = 1\,\mu$m and $E_0=2.3$\,V/nm, we measure an average current of $J_\text{avg}=50$\,pA at 80\,MHz repetition rate, when the CEP is set to $[0,\pi]$ (Fig.~2\textbf{c}). We note that for $L=1$\,$\mu$m the electric field strength across the graphene strip is approximately constant (\fig{Fig_S4}\textbf{b}) and, thus, this case most appropriate for a comparison. With identical optical parameters, in the simulation we obtain $Q(t\rightarrow\infty)\approx-0.04$\,e per pulse (equal to $J_\text{avg}=0.51$\,pA at 80\,MHz repetition rate) for a GNR with a width of 5.2\,nm and independently of its length (Fig.~3\textbf{b}). Utilizing a linear scaling of the transferred charge as a function of the graphene strip width and the coupling $\Gamma$ and taking into account focal averaging over the intensity profile given in the experiment, the simulation matches perfectly the experiment for $\Gamma=0.4$\,eV. Therefore, we conclude that $\Gamma=0.1$\,eV as used in the simulations shown in Fig.~3 assumes a reasonable quantitative coupling.

\subsubsection*{Quantum dynamics.}
The Hamiltonian of the composite system (Eq. \ref{eq:composite_hamiltonian}) is an effective single-particle Hamiltonian of the form $H(t)=\sum_{\nu\mu}h_{\nu\mu}(t)\hat{a}^{\dagger}_{\nu}\hat{a}_{\mu}$. As such, its electronic properties are completely determined by the single-particle reduced density matrix $
\rho_{\nu\mu}(t)=\langle \hat{a}^{\dagger}_{\nu} \hat{a}_{\mu}\rangle$. The laser-induced dynamics of the GNR in the junction was obtained by solving the Liouville von Neumann equation for $\rho_{\nu\mu}$ using the non-equilibrium Green's function method (NEGF) developed by Chen \emph{et al.}~\cite{Zhang2013,Zheng2007}. In this method, in the junction region $\rho_{\nu\mu}(t)$ satisfies
\begin{equation}
	\label{eq:liouville_eq}
	i\hbar \frac{\mathrm{d}}{\mathrm{d}t}\rho_{\nu\mu}(t)=\langle [\hat{a}^{\dagger}_{\nu} \hat{a}_{\mu},H(t)]\rangle-\sum_{\alpha}(\varphi_{\alpha}(t)-\varphi_{\alpha}^{\dagger}(t)), 
\end{equation} 
where the first term quantifies the unitary dynamics while the $\varphi_{\alpha}(t)$ and $\varphi_{\alpha}^{\dagger}(t)$ are auxiliary density matrices that incorporate charge injection and subtraction by the metallic contacts into the GNR. Chen and co-workers~\cite{Zheng2007} developed a computational efficient set of equations (Eqs. (3), (12) and (14) in Ref.~\cite{Zheng2007}) to capture time-dependent transport by invoking the WBL and a Pad\'e expansion of the Fermi-Dirac distribution function. The Pad\'e expansion allows for analytically solving the energy integrals that appear in the definition of the self-energies. Here, we use 50~Pad\'e functions for the expansion for representing the metal contacts, a time step of $\Delta t=0.003$\,fs for the integration using the Runge-Kutta method of order four, and a Fermi energy $\mu_\text{F}=0$ right at the Dirac cone. All numerical parameters were checked for convergence.

\begin{figure*}[t!]
	\begin{center}
		\includegraphics[width=14cm]{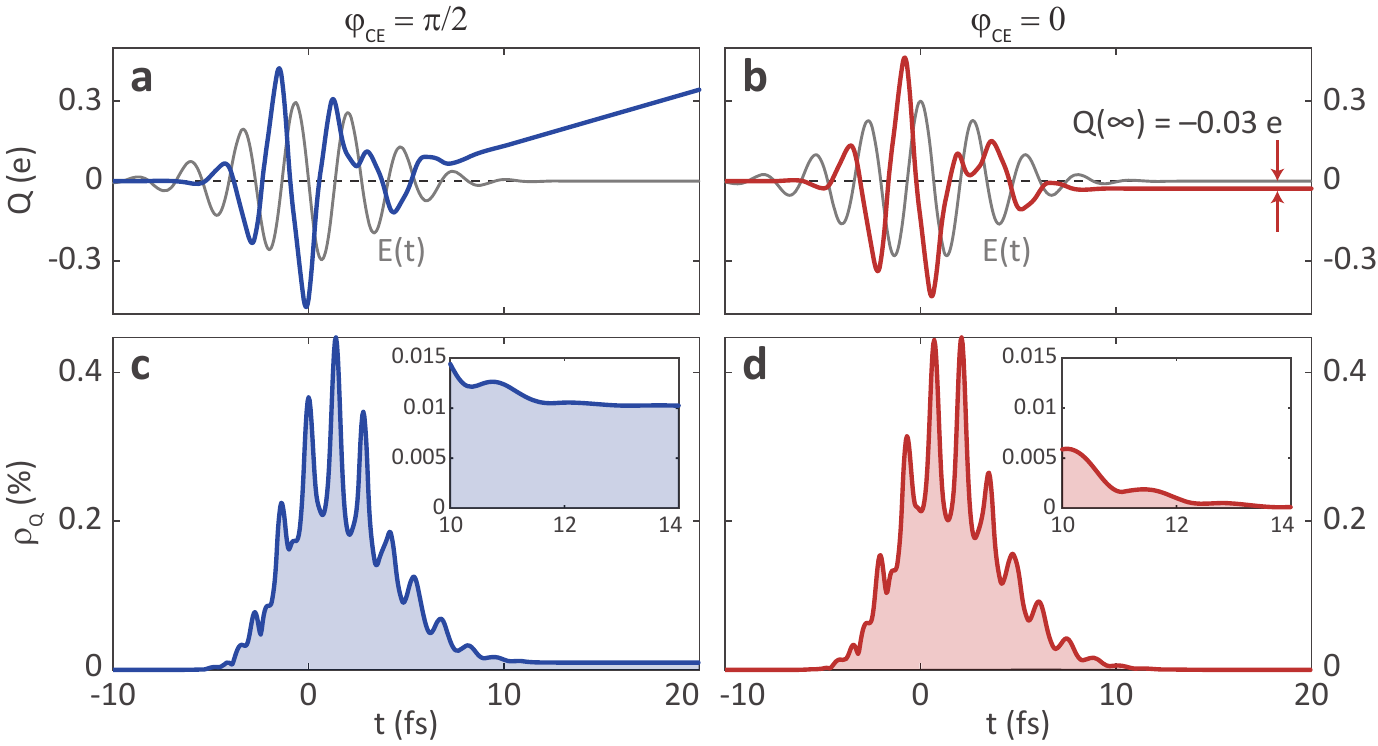}
		\caption{\textbf{Transient evolution of charge motion and electron population obtained from a TDSE model.} \textbf{a, b,} Charge motion for a CEP of  $\varphi_\text{CE}=\pi/2$ (\textbf{a}) and $\varphi_\text{CE}=0$ (\textbf{b}). The gray lines show the electric field of the pulses with $E_0=2.3$\,V/nm, centered at $t=0$. \textbf{c, d,} Normalized electron population $\rho_Q$ contributing to the charge transfer. The population is normalized to the number of available states in the first Brillouin zone of graphene. For $\varphi_\text{CE}=0$, $\rho_Q$ returns to zero after the pulse is gone, while for $\varphi_\text{CE}=\pi/2$, a residual population remains and leads to an increasing charge transfer, see also insets.}
		\label{Fig_S7}
	\end{center}
\end{figure*}

Numerically capturing the experimental dynamics by solving the Liouville von Neuman equation in real space requires using long GNR models with $10^3-10^5$ atoms. To keep the computations numerically tractable but still capture an experimentally relevant density of states, we consider the Hilbert space of the GNR in the $0.2-1.3$\,eV energy range only. This region captures most of the Dirac cone where the light-driven dynamics is expected to occur. This truncation was performed by unitary transforming the GNR Hamiltonian and the coupling to the leads to the energy eigen-basis of the pristine GNR. This Hilbert space truncation influences the magnitude of the effect but leaves the basic qualitative features of the dynamics intact. 

To phenomenologically incorporate dephasing effects in the dynamics, we express \eq{eq:liouville_eq} in the single-particle energy eigenbasis of pristine graphene $\{ |\epsilon\rangle\}$
\begin{equation}
	\begin{split}
		\label{eq:liouville_eq2}
		i\hslash \frac{d}{dt}\rho_{\epsilon\epsilon'}(t)&=\langle [\hat{a}^{\dagger}_{\epsilon} \hat{a}_{\epsilon'},H(t)]\rangle\\
		&-\sum_{\alpha}(\varphi_{\alpha}(t)-\varphi_{\alpha}^{\dagger}(t))\\
		&- i \frac{\hslash}{T_2}\rho_{\epsilon\epsilon'}(t)(1- \delta_{\epsilon, \epsilon'}),
	\end{split}
\end{equation} 
and introduce terms that dynamically force the off-diagonal elements of the density matrix $|\rho_{\epsilon, \epsilon'}(t)|^2$ ($\epsilon\ne \epsilon'$) to decay in a characteristic timescale $T_2$. Dephasing may be introduced by electron-electron or phonon interaction as well as scattering at lattice defects that emerge typically on 10 to 100\,fs~\cite{Lui2010,Gierz2013,Vampa2014,Floss2019}. For dephasing in graphene, a lower boundary of 22\,fs was obtained from two-color current injection~\cite{Heide2021b}. Based on these results, we choose $T_2 = 20$\,fs.

The average current flowing through the GNR is 
$I(t)=(I_{\text{R}}(t)-I_{\text{L}}(t))/2$
where $I_{\alpha}(t)=-e\frac{\mathrm{d}}{\mathrm{d}t}\sum_q\langle\hat{a}^{\dagger}_{\alpha,q}\hat{a}_{\alpha,q}\rangle$ is the current entering into lead $\alpha$.  A phenomenological relaxation of the current $I(t>0)$ is introduced by an exponential decay. It accounts for momentum relaxation through carrier thermalization ($\sim$50\,fs) \cite{Malic2011,Song2013,Brida2013}, and phonon and defect scattering ($\sim$26\,fs, see ``Sample fabrication and characterization'') restricting ballistic transport to few 10s of fs. Subsequent diffusive charge transport of ballistically displaced carriers is limited by cooling via carrier-phonon interaction with optical phonons ($\sim$100\,fs) and acoustic phonons ($\sim$0.7\,ps)~\cite{Lui2010,Gierz2013}. Following these time scales, we apply a 100-fs exponential decay to the current. Laser-induced symmetry breaking is monitored by quantifying the net transferred charge at a given time $t$ 
\begin{equation}
	\label{eq:charge}
	Q(t)=\int_{-\infty}^{t}I(t')\mathrm{d}t'
\end{equation}

The simulations are performed at the Center for Integrated Research Computing at the University of Rochester where they require a computing time of up to 8 days per trace. Larger graphene structures, as used in the experiments, are beyond state-of-the-art resources as the computational effort increases exponentially with system size.

\subsubsection*{Population dynamics.}

To isolate the distinct role of real and virtual charge carriers to the charges obtained from our simulation, we simplify the above model system to the tight-binding Hamiltonian $H_S(t)$ (\eq{eq:graphenehamil}) only~\cite{Ishikawa2010, Kelardeh2015}. The graphene is modeled by two bands representing the valence band (VB) $\varepsilon_+(\textbf{k})$ and the conduction band (CB) $\varepsilon_-(\textbf{k})$. We model the light-matter interaction by solving the time-dependent Schrödinger equation (TDSE) in length gauge. For the optical field, we use Gaussian pulses as sown in \fig{Fig_S6}. The intraband current is computed as
\begin{equation}
	\label{eq:j_intra}
	j^\text{intra}(t)=e\sum_{m=\text{CB,VB}}\int_\text{BZ}v^{(m)}_\textbf{k}\rho^{(m)}_{\textbf{k}_0}(t)\frac{\mathrm{d}\textbf{k}}{(2\pi)^2},
\end{equation}
with the electron velocity $v^{(m)}_\textbf{k}=\hbar^{-1}\frac{\partial\varepsilon_\pm(\textbf{k})}{\partial k_x}$ and the electron population $\rho^{(m)}_\textbf{k}(t)$. The integral is taken over the Brillouin zone (BZ). The transferred charge $Q(t)$ is obtained according to Eq.~\eqref{eq:charge}.

In \fig{Fig_S7} the charge transfer is shown for $\varphi_\text{CE}=\pi/2$ (panel \textbf{a}) and $\varphi_\text{CE}=0$ (panel \textbf{b}). Both curves show good qualitative agreement with the results obtained from the full real-space simulation (Fig.~3) and reproduce the crucial CEP dependence:
The transient charge oscillations appear and are even more pronounced, as they are not damped by the electrode coupling $\Gamma$. For $\varphi_\text{CE}=\pi/2$ (\fig{Fig_S7}\textbf{a}) the slope of $Q$, reflects the momentum of ballistically launched charge carriers. However, here the slope is one order of magnitude larger compared to Fig.~3\textbf{a}, as only the graphene is considered in the model; consequently, the initial momentum is unimpaired by dephasing, electrode coupling and interfacial reflections and augmented by the bulk nature of this graphene model. In good quantitative agreement with the full simulation (Fig.~3\textbf{b}), we obtain a net charge displacement of $Q(t\rightarrow\infty)=-0.03\,e$ for $\varphi_\text{CE}=0$ after the pulse is gone (\fig{Fig_S7}\textbf{b}). When the translational symmetry of the graphene lattice is broken by electrodes attached to the graphene, the net charge displacement can be probed as a current (Fig.~3\textbf{b}).

Any population imbalance is identified by taking the population difference driven by fields of opposite CEP, $\Delta\rho_{\textbf{k}_0}^{(m)}(t)=\rho_{\textbf{k}_0}^{(m)}(t,\varphi)-\rho_{\textbf{k}_0}^{(m)}(t,\varphi+\pi)$ ($m=\text{CB, VB}$). We then define the normalized population that effectively contributes to a charge transfer by weighting it with the absolute of $\Delta\rho_{\textbf{k}_0}^{(m)}(t)$ and $v^{(m)}_\textbf{k}$:
\begin{equation}
	\label{eq:effective_charge_carrier_pop}
	\rho_Q(t)=\sum_{m=\text{CB,VB}}\frac{\int_\text{BZ}\lvert v_\textbf{k}^{(m)}\cdot\Delta\rho_{\textbf{k}_0}^{(m)}(t)\rvert\,\rho_{\textbf{k}_0}^{(m)}(t,\varphi)\mathrm{d}\textbf{k}}{\int_\text{BZ}\lvert v^{(m)}_\textbf{k}\rvert \mathrm{d}\textbf{k}}.
\end{equation}
The denominator normalizes $\rho_Q$ by the total number of states available in the Brillouin zone, again weighted with the electron velocity $v_\textbf{k}^{(m)}$.

Figures~\ref{Fig_S7}\textbf{c} and \textbf{d} show the normalized population $\rho_Q$ as a function of time. Clearly, for $\varphi_\text{CE}=0$, $\rho_Q$ manifests itself to be virtual only as it returns completely back to zero after the pulse is gone (\fig{Fig_S7}\textbf{d}). Also for $\varphi_\text{CE}=\pi/2$ substantial virtual population is excited (\fig{Fig_S7}\textbf{c}). However, due to the symmetry of the driving field no net polarization is induced by this virtual population, see also Fig.~3\textbf{b}. In turn, the linear increase of $Q$ after the pulse, shown in \fig{Fig_S7}\textbf{a}, is due to a residual and, therefore, real population of conduction band states (\fig{Fig_S7}\textbf{c}, inset).

\subsubsection*{Two-pulse scheme and reconstruction of logic gates.}

\begin{figure}[b!]
	\begin{center}
		\includegraphics[width=7.5cm]{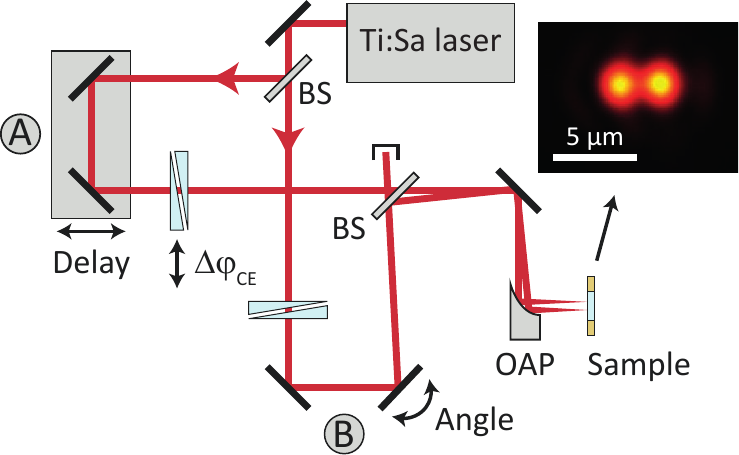}
		\caption{\textbf{Two-pulse experiment.} CEP-stable laser pulses from a Ti:Sa oscillator are split into pulse copies ($A$, $B$) in a Michelson interferometer. The SiO$_2$ wedge pairs are used to balance the dispersion (path $A$ and $B$), and to vary $\Delta\varphi_\text{CE}$ (path $A$). The temporal delay is changed by variation of path length $A$. Introducing an angle in path $B$ results in spatially separated foci on the sample, see microscope image. BS, beam splitter; OAP, off-axis parabolic mirror.}
		\label{Fig_S8}
	\end{center}
\end{figure}

\begin{figure*}
	\begin{center}
		\includegraphics[width=13.5cm]{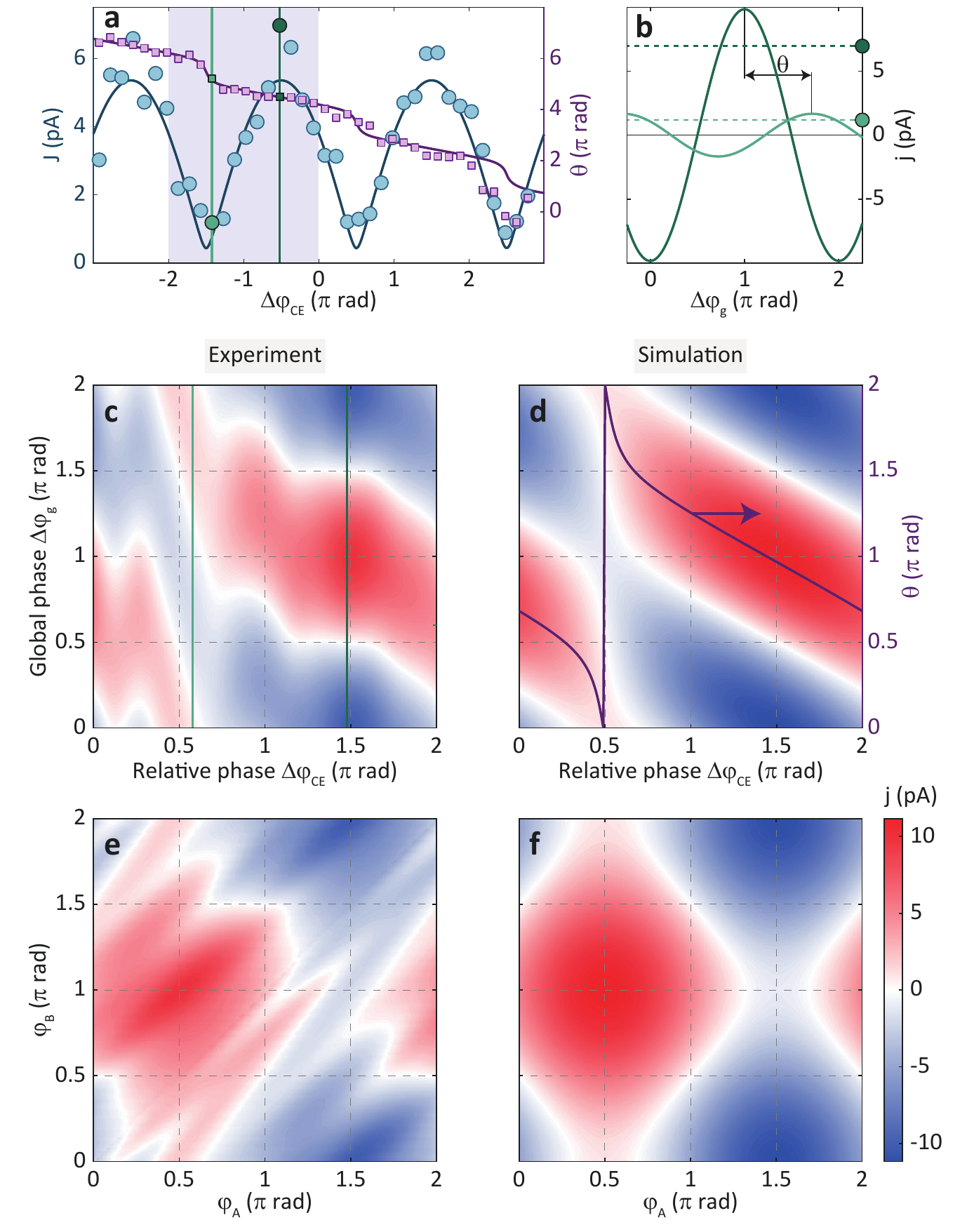}
		\caption{\textbf{CEP dependence of currents for logic switching.} 
			\textbf{a,} Measured and simulated CEP-dependent current (blue, data and simulation as in Fig.~4\textbf{b}) and lock-in phase (purple). The shaded range $\Delta\varphi_\text{CE}=[-2\pi,0]$ is further analyzed in (\textbf{c}), where it is mapped to $\Delta\varphi_\text{CE}=[0,2\pi]$. \textbf{b,} Using \eq{eq:current_demodulated} of Methods, the $\Delta\varphi_\text{g}$ dependence of the current is exemplified for two data points (light and dark green) of (\textbf{a}). Dashed lines indicate the root mean square values, i.e., the currents shown in (\textbf{a}). \textbf{c,} Measured current as a function of $\Delta\varphi_\text{CE}$ and demodulated along $\Delta\varphi_\text{g}$. The light and dark green lines correspond to the data points marked in (\textbf{a}) and analyzed in (\textbf{b}). \textbf{d,} Simulated current as a function of $\Delta\varphi_\text{CE}$ and $\Delta\varphi_\text{g}$. The model curve of $\theta$ (purple line) is indicated with an arbitrary offset with respect to (\textbf{a}). \textbf{e,} Measured current (\textbf{c}) in the basis of individual CEPs $\varphi_A$ and $\varphi_B$. \textbf{f,} Simulated current in the same basis as (\textbf{e}).}
		\label{Fig_S9}
	\end{center}
\end{figure*}

\begin{figure*}
	\begin{center}
		\includegraphics[width=13.5cm]{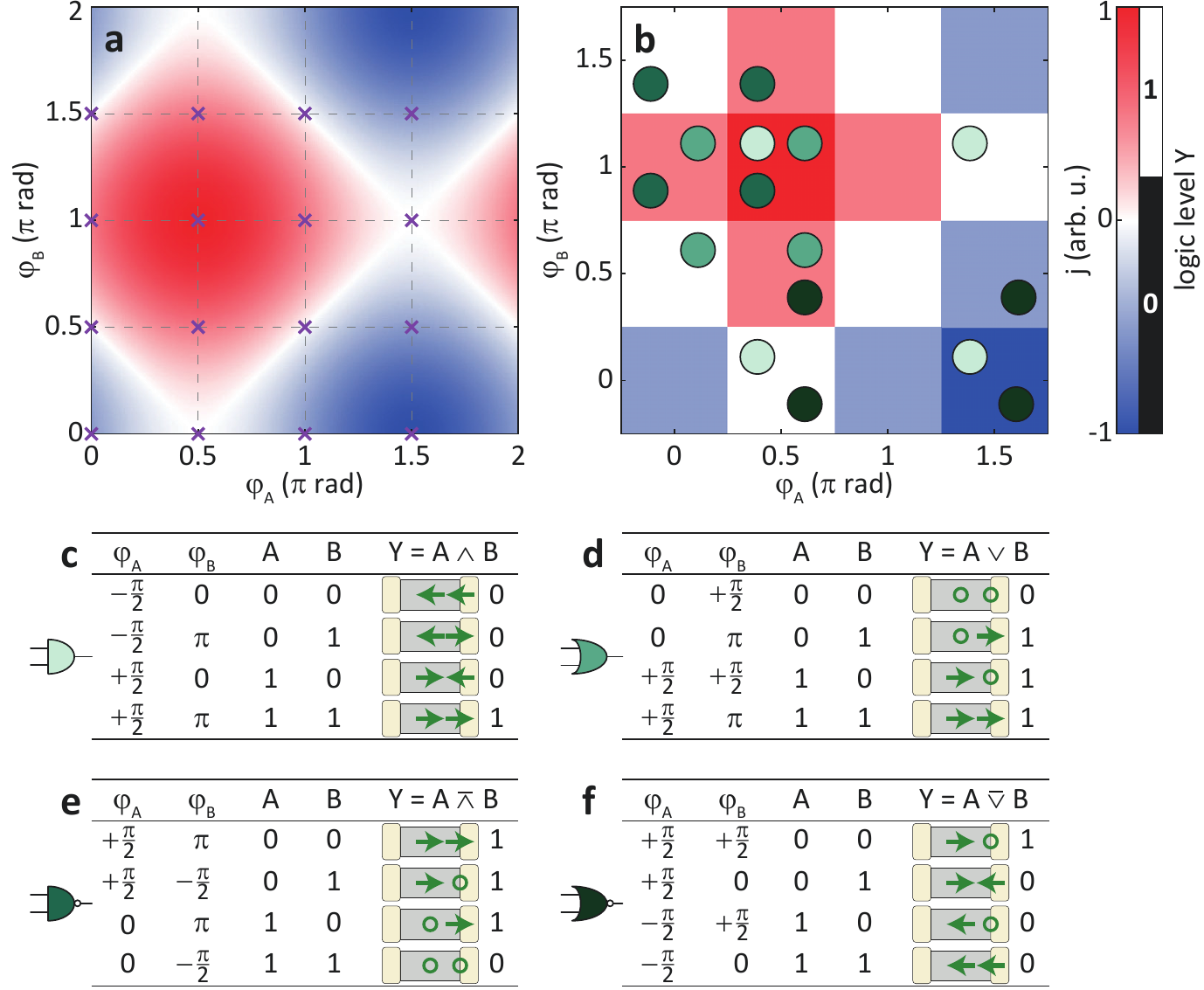}
		\caption{\textbf{Realization of various logic gates.} \textbf{a,} Simulated CEP-dependent current as a function of $\varphi_A$ and $\varphi_B$ with equal current amplitudes $J_A$, $J_B$ induced by laser pulses $A$ (bulk graphene) and $B$ (interface). The purple crosses mark phases $\varphi_A$, $\varphi_B$ considered in (\textbf{b}). \textbf{b,} CEP-dependent current for 16 combinations of $\varphi_A$ and $\varphi_B$ as needed for the operation of logic gates. Positive current is assigned to an output logic level $Y$ of 1 while zero or negative current is assigned to 0 (see color bar). The green dots mark the combinations required for the formation of the logic gates shown below. \textbf{c--f,} Truth tables of AND (\textbf{c}), OR (\textbf{d}), NAND (\textbf{e}) and NOR (\textbf{f}) are obtained by appropriate choice of CEPs $\varphi_A$ and $\varphi_B$ as marked in (\textbf{b}). The green arrows (rings) on the insets depicting the heterostructure mark the current direction (zero current) driven on the graphene and at the interface.}
		\label{Fig_S10}
	\end{center}
\end{figure*}

To generate two pulsed laser beams, the incoming beam is fed into a dispersion-balanced Michelson interferometer where two pulsed beam copies ($A$, $B$) are generated (\fig{Fig_S8}). A pair of SiO$_2$ wedges in the interferometer arm $A$ is used to control the relative CEP between the two laser pulses. The end mirror of interferometer arm $B$ is detuned such that both laser pulses leave the interferometer collinearly up to a small angle. Subsequent focusing of the laser pulses with an off-axis parabolic mirror results in a separation of the foci of pulse $A$ and $B$ by 2.5\,$\mu$m (\fig{Fig_S8}, microscope image). Pulse $A$ is placed on the center of a $5\times1.8$\,$\mu$m$^2$ graphene strip to drive a ballistic current by real charge carriers only, while pulse $B$ illuminates one of the gold-graphene interfaces, where current is injected predominantly by virtual charge carriers. A temporal delay of approximately 85\,fs ($A$ before $B$) is introduced by the interferometer to deploy a ballistic current by pulse $A$ reasonably before it is transiently switched by one (or more) pulses $B$. We note that, here, currents were measured with an integration time of 7\,s, therefore the observed current is insensitive to the temporal delay between laser pulses $A$ and $B$.

To construct a logic gate with logic levels of the inputs encoded in the CEP of the two laser pulses we evaluate the residual current for all possible combinations of $\varphi_A$ and $\varphi_B$. As described above, we measure CEP-dependent current amplitudes $J$ and their phases $\theta$ in a dual-phase lock-in scheme referenced to the carrier-envelope-offset frequency ($f_\text{CEO}$) that imprints a global periodic CEP modulation $\Delta\varphi_\text{g}=2\pi f_\text{CEO}\Delta t$ to both laser pulses. By additionally controlling the relative CEP $\Delta\varphi_\text{CE}$ by means of a SiO$_2$ wedge pair in interferometer arm $A$, we can realize arbitrary CEPs $\varphi_A$ and $\varphi_B$ and measure the resulting current. Note that $\theta$ is no longer directly proportional to a single CEP as two pulses with different CEPs drive currents based on real and virtual charge carriers that have an intrinsically different CEP dependence (see Supplementary Video).

Here we unfold the full $\varphi_A$, $\varphi_B$ dependence of the data shown in Fig.~4\textbf{b} (see also \fig{Fig_S9}\textbf{a} with $\theta$ added). Each data point $J$ represents the root mean square value of an alternating current $j$, modulated at $f_\text{CEO}$. Hence, to obtain this alternating current as a function of $\Delta\varphi_\text{g}$, each data point is multiplied with a sine function shifted by the respective lock-in phase $\theta$ and scaled by a factor of $\sqrt{2}$ to obtain the peak values
\begin{equation}
	\label{eq:current_demodulated}
	j(\Delta\varphi_\text{CE},\Delta\varphi_\text{g})=\sqrt{2}J(\Delta\varphi_\text{CE})\cdot\sin(\Delta\varphi_\text{g}-\theta(\Delta\varphi_\text{CE})),
\end{equation}
representing the demodulation of the dual-phase lock-in amplifier output signal. In \fig{Fig_S9}\textbf{b} we exemplify the $\Delta\varphi_\text{g}$ dependence of two data points, namely the minimum and maximum current (\fig{Fig_S9}\textbf{a} and \textbf{b}, light and dark green data points). The phase of the sine functions is given by their respective $\theta$ (\fig{Fig_S9}\textbf{a}, green squares), while their root mean square values (dashed lines) yield the amplitudes shown in \fig{Fig_S9}\textbf{a} (green circles). 

The demodulation of all data points within one period of $\Delta\varphi_\text{CE}$ (\fig{Fig_S9}\textbf{a}, shaded range) according to \eq{eq:current_demodulated} yields the full $\Delta\varphi_\text{CE}$ and $\Delta\varphi_\text{g}$ dependence of the current. This is shown in \fig{Fig_S9}\textbf{c}, where the color code indicates a sine function for each column, similar to \fig{Fig_S9}\textbf{b}. Note here that for $\Delta\varphi_\text{CE}=+\pi/2$ the current remains close to zero regardless of the values of the individual CEPs $\varphi_A$ and $\varphi_B$.

We can then map this current to a basis spanned by $\varphi_A$ and $\varphi_B$ according to
\begin{equation}
	\begin{split}
		\label{eq:current_basis_trafo}
		\tilde{j}\left(\varphi_A,\varphi_B\right)&=\\
		j(\Delta\varphi_\text{CE}&=\left(\varphi_A-\varphi_B\right)\bmod2\pi,\Delta\varphi_\text{g}=\varphi_B),
	\end{split}
\end{equation}
as shown in \fig{Fig_S9}\textbf{e}. 

To support the observed dependencies of the total current on $\varphi_A$ and $\varphi_B$, we apply a simple model where we assume $j_A(\varphi_A)=J_A\cdot\sin(\varphi_A)$ and $j_B(\varphi_B)=-J_B\cdot\cos(\varphi_B)$ for currents driven by laser pulse $A$ and $B$, respectively. The sine and a minus cosine functions are chosen for $j_A$ and $j_B$, respectively, to reflect the proper CEP dependence of currents injected by real and virtual charge carriers. The current amplitudes $J_A$ and $J_B$ are evaluated from measurements on the identical heterostructure with laser pulse $A$ on graphene only ($J_A=6.1$\,pA) and laser pulse $B$ on the interface only ($J_B=5.1$\,pA).

The total current $\tilde{j}(\varphi_A,\varphi_B)=j_A+j_B$ under illumination with both laser pulses is shown in \fig{Fig_S9}\textbf{f}, where the axes are spanned by $\varphi_A$ and $\varphi_B$, like in \fig{Fig_S9}\textbf{e}. A reverse transformation to Eq.~\eqref{eq:current_basis_trafo},
\begin{equation}
	\begin{split}
		\label{eq:current_basis_backtrafo}
		j\left(\Delta\varphi_\text{CE},\Delta\varphi_\text{g}\right)&=\\
		\tilde{j}(\varphi_A=(\Delta&\varphi_\text{CE}+\Delta\varphi_\text{g})\bmod2\pi,\varphi_B=\Delta\varphi_\text{g}),
	\end{split}
\end{equation}
yields the simulated current in the same basis as \fig{Fig_S9}\textbf{c}, see \fig{Fig_S9}\textbf{d}. The simulation curves shown in \fig{Fig_S9}\textbf{a} (and Fig.~4\textbf{b}) can be extracted from this panel (see also the Supplementary Video): The simulated lock-in phase tracks the current crest as a function of $\Delta\varphi_\text{CE}$ (\fig{Fig_S9}\textbf{d}, purple line). The root mean square value of each column gives the simulation curve for the current measured by the dual-phase lock-in amplifier (Figs.~\ref{Fig_S9}\textbf{a} and 4\textbf{b}, blue line). It is scaled by a factor of 0.68 to fit best to the experimental data. 

In both representations shown in \fig{Fig_S9}\textbf{c}--\textbf{f} our simple model results exhibit good agreement with the experimentally obtained currents: both current amplitude and direction fit well. Resting on this, we may use our model as a basis for illustrating the formation of logic gates. We simplify the current map of \fig{Fig_S9}\textbf{f} by choosing arbitrary equal amplitudes $J_A$, $J_B$, see \fig{Fig_S10}\textbf{a}. Since binary inputs of the CEP are required for logic levels, we reduce this map to integer multiples of $\pi/2$, yielding a table of 16 possible combinations of $\varphi_A$ and $\varphi_B$ (\fig{Fig_S10}\textbf{b}). We assign the resulting currents to logic output levels of 1 for positive current and 0 for zero or negative current. The gate types AND, OR, NAND and NOR shown in \fig{Fig_S10}\textbf{c}--\textbf{f} can be located straight-forwardly on the table of \fig{Fig_S10}\textbf{b} (see green dots). Note that for each of the gates the assignment of CEPs $\varphi_A$, $\varphi_B$ to logic input levels 0 or 1 may be different, as shown in the truth tables. Also note that the NOR gate is already functionally complete, thus allowing one to build an entire processor using NOR gates only.

\end{document}